\documentclass{article}

\usepackage{amsmath,amsfonts,amsthm}
\usepackage{graphicx}
\usepackage{fixmath}
\usepackage{amssymb,latexsym}
\usepackage{mathrsfs,amsbsy}
\usepackage{dsfont}
\usepackage{algorithm}
\usepackage{eucal}
\usepackage{colortbl}
\usepackage{multirow}
\usepackage[end]{algpseudocode}
\usepackage{float}
\graphicspath{{./figures/}}
\usepackage{booktabs}
\usepackage{caption}
\usepackage{subfigure}
\usepackage{microtype}

\theoremstyle{definition}

\theoremstyle{remark}

\numberwithin{equation}{section}
\numberwithin{subsection}{section}

\title{
Uniqueness of Transformations with Prescribed Jacobian Determinant and Curl-Vector
}
\author{Zicong Zhou\\
University of Texas at Arlington\\
zicong.zhou@mavs.uta.edu
\and 
Xi Chen\\
xi.chen@mavs.uta.edu
\and 
Xian Xin Cai\\
newnewice@sina.com
\and 
Guojun Liao\\
University of Texas at Arlington\\
liao@uta.edu
\date{December 9, 2017}
}

\begin{document}
\maketitle    
\begin{abstract}
Numerical examples demonstrated that a prescribed positive Jacobian determinant alone can not uniquely determine a diffeomorphism. It is conjectured that the uniqueness of a transformation can be assured by its Jacobian determinant and the curl-vector. In this work, we study the uniqueness problem analytically and propose an approach to the proof of the uniqueness of a transformation with prescribed Jacobian determinant and curl-vector.
\end{abstract}
\section{Introduction}
In the research area of numerical mesh generation, our group had developed the $deformation$ $method$ \cite{Ands} \cite{Cai} and the $variational$ $method$ \cite{XiChen} \cite{XiChen2}. Both of these two methods were formulated based on the construction of diffeomorphisms between domains in $\mathbb R^n$ \cite{Dac:Mos}\cite{Liao}. The former is well studied and understood. Its theoretical framework and applications had been showed in \cite{Cai} \cite{Liu} and it had been furtherer established to generate higher order mesh \cite{Zhou}. 

Construction of diffeomorphism is an interesting topic. In \cite{Liao}, we proposed the problem of generating a diffeomorphism with prescribed Jacobian deteminant and the cur-vector. Indeed, we formulated the $variational$ $method$, which numerically constructs diffeomorphisms with prescribed positive Jacobian determinant and the curl-vector in $L^{2}-norm$. The following example demonstrates an important fact about the curl-vector of a transformation.

\begin{center}
$\mathbf{Example}$ 1: Effect of the Curl-Vector
\end{center}
{\setlength{\parindent}{0cm}
The Jacobian determinant can not uniquely determine a transformation without the curl-vector. The following two meshes are generated based on the intensity of a Mona Lisa's portrait. They have the same Jacobian determinant but with different curl-vectors.
\begin{figure}[H]
	\caption*{$\mathbf{1st}$ mesh generated by the $deformation$ $method$}
	{\includegraphics[width=12cm,height=12cm]{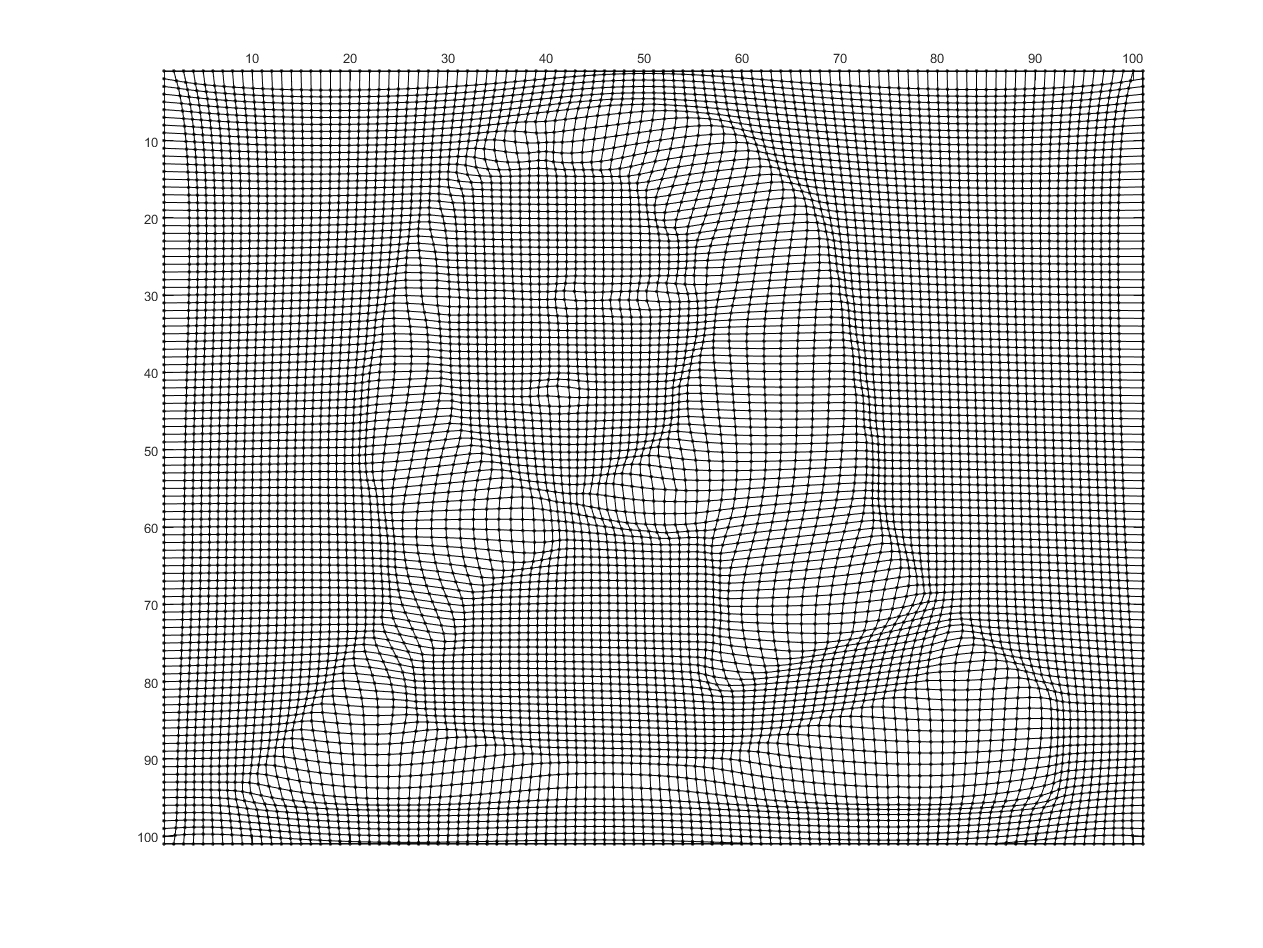}}
\end{figure}
 	
\begin{figure}[H]
	\caption*{$\mathbf{2nd}$ mesh reconstructed by the $variational$ $method$}
	{\includegraphics[width=12cm,height=12cm]{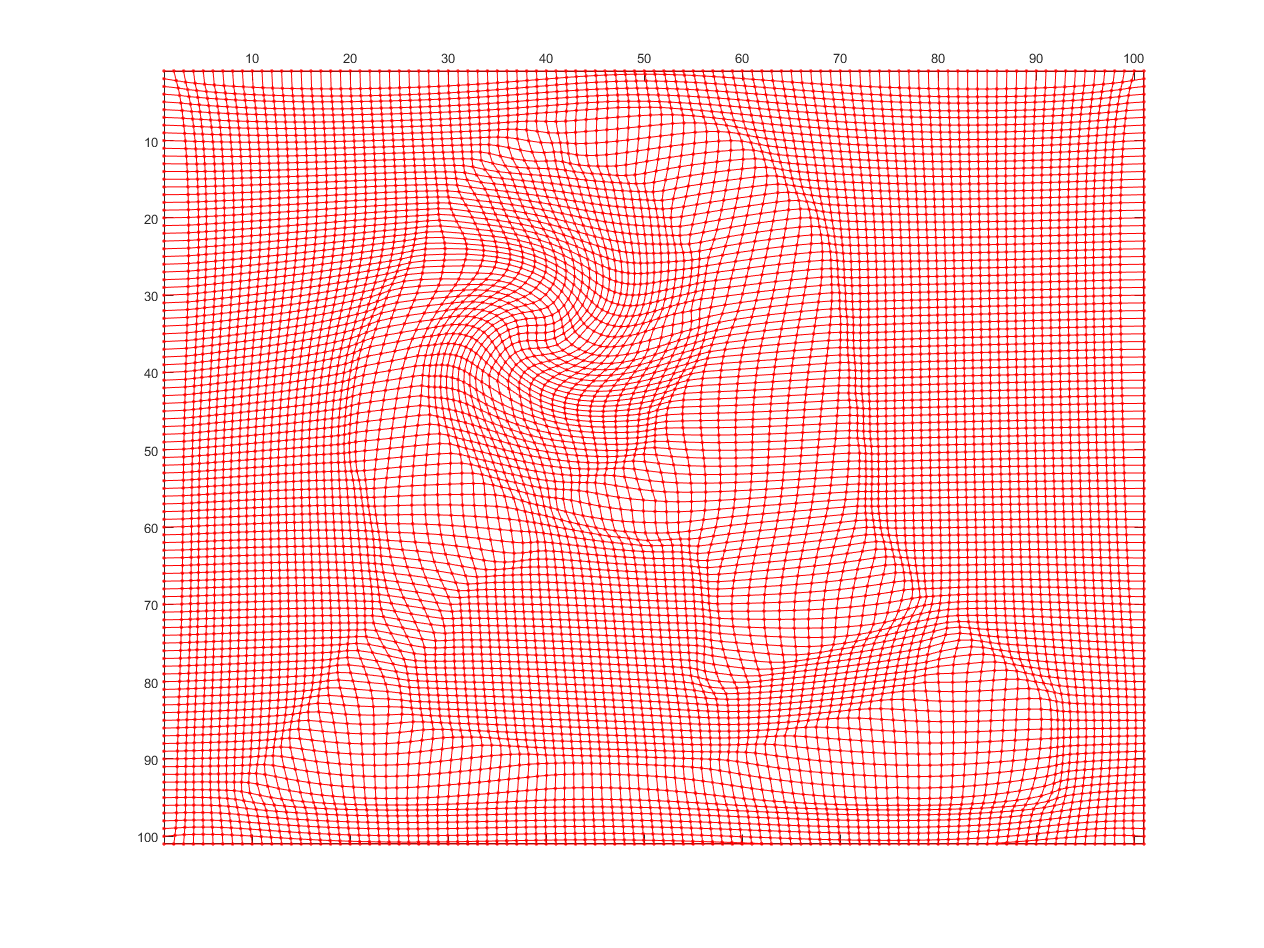}}
\end{figure}
}
Both of these meshes have the same distribution of cell-size, which approximates the Jacobian determinant. But the second has different curl-vector from the first mesh on the face, where the curl-vector can be understood as the rotation of grid lines. As it can be seen, on the facial part, the grid lines of second mesh appear rotated, while the first has horizontal and vertical grid lines there.
\begin{center}
$\mathbf{Example}$ 2 Recovering $\pmb{T_{0}}$ from $\text{det}\nabla(\pmb{T_{0}})$ and $\text{curl}(\pmb{T_{0}})$
\end{center}
{\setlength{\parindent}{0cm}
Given a transformation $\pmb{T_{0}}$ from the uniform Cartesian mesh on a square onto itself, we calculate the Jacobian determinant $\text{det}\nabla(\pmb{T_{0}})$ and the $\text{curl}(\pmb{T_{0}})$. Then we use the $variational$ $method$ to reconstruct $\pmb{T_{0}}$, which will be reviewed below. The recovered transformation $\pmb{T_{1}}$ is only based on $\text{det}\nabla(\pmb{T_{0}})$ as prescribed Jacobian determinant and is shown in the following Figure (a), where the nodes of $\pmb{T_{0}}$ are marked by the stars and the mesh lines are marked by black lines. The mesh lines of the reconstructed $\pmb{T_{1}}$ are marked by red lines. In Figure (b) and (c), the enlarged views of the two rectangular regions are shown.  
\begin{figure}[H]
	\subfigure[$\pmb{T_{0}}$ --- black(w/ *) and $\pmb{T_{1}}$ --- red]{{\includegraphics[width=10cm,height=9cm]{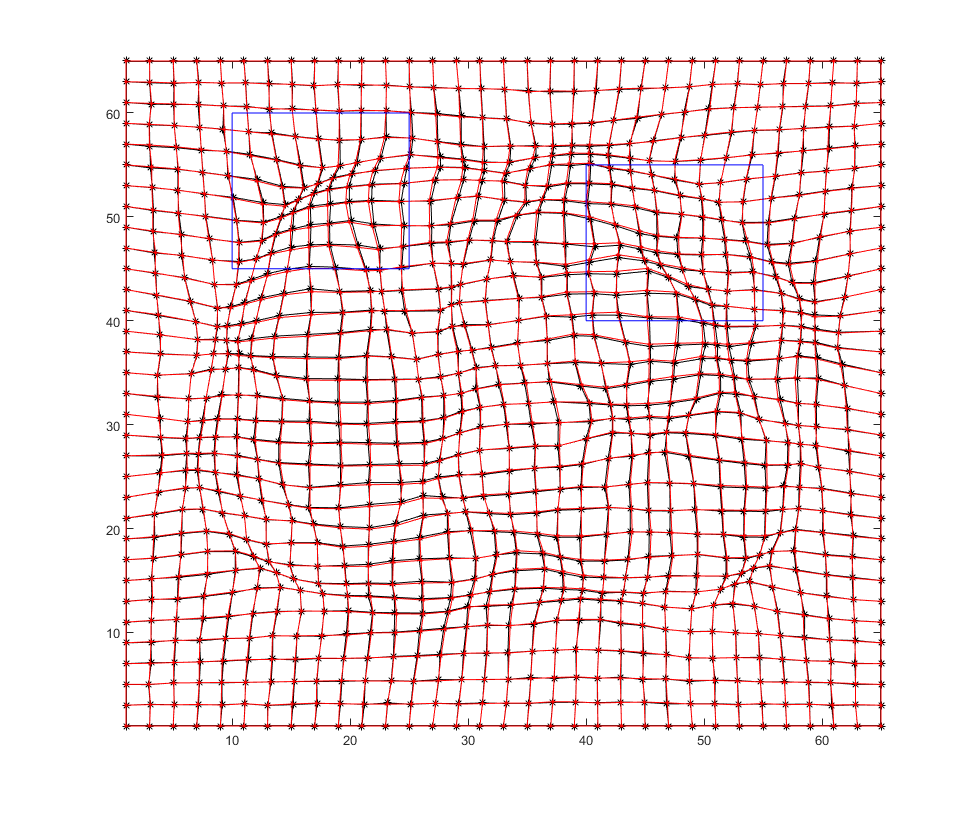}}}\\
	\subfigure[Enlarged view of the left rectangle]{{\includegraphics[width=6cm,height=5.4cm]{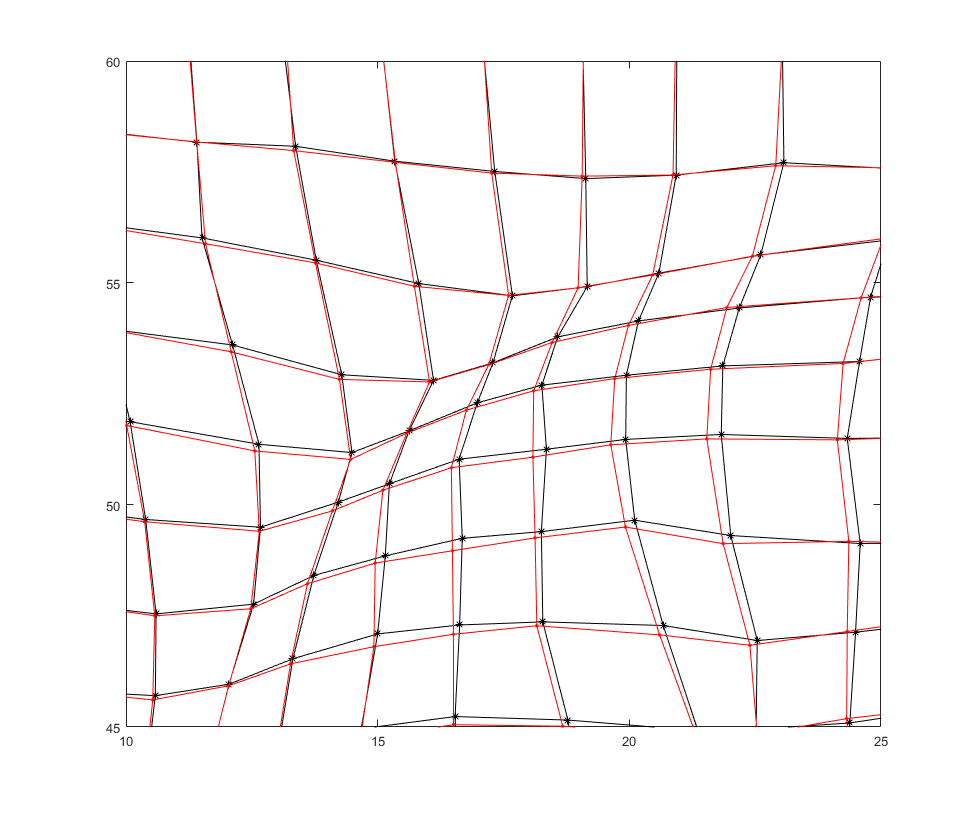}}}
	\subfigure[Enlarged view of the right rectangle]{{\includegraphics[width=6cm,height=5.4cm]{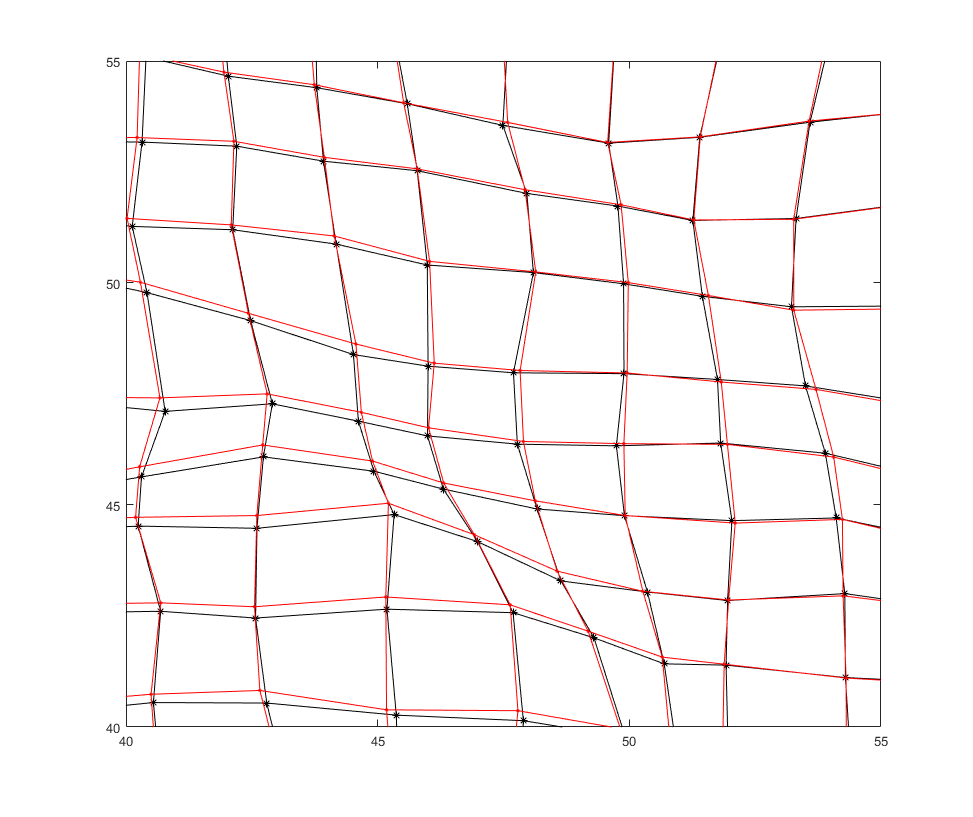}}}
\end{figure}
As it can be seen, the red lines do not perfectly overlap with the black lines. Next, we use both the $\text{det}\nabla(\pmb{T_{0}})$ and the $\text{curl}(\pmb{T_{0}})$ to reconstruct the transformation $\pmb{T_{2}}$. 
\begin{figure}[H]	
	\subfigure[$\pmb{T_{0}}$ --- black(w/ *) and $\pmb{T_{2}}$ --- red]{{\includegraphics[width=10cm,height=9cm]{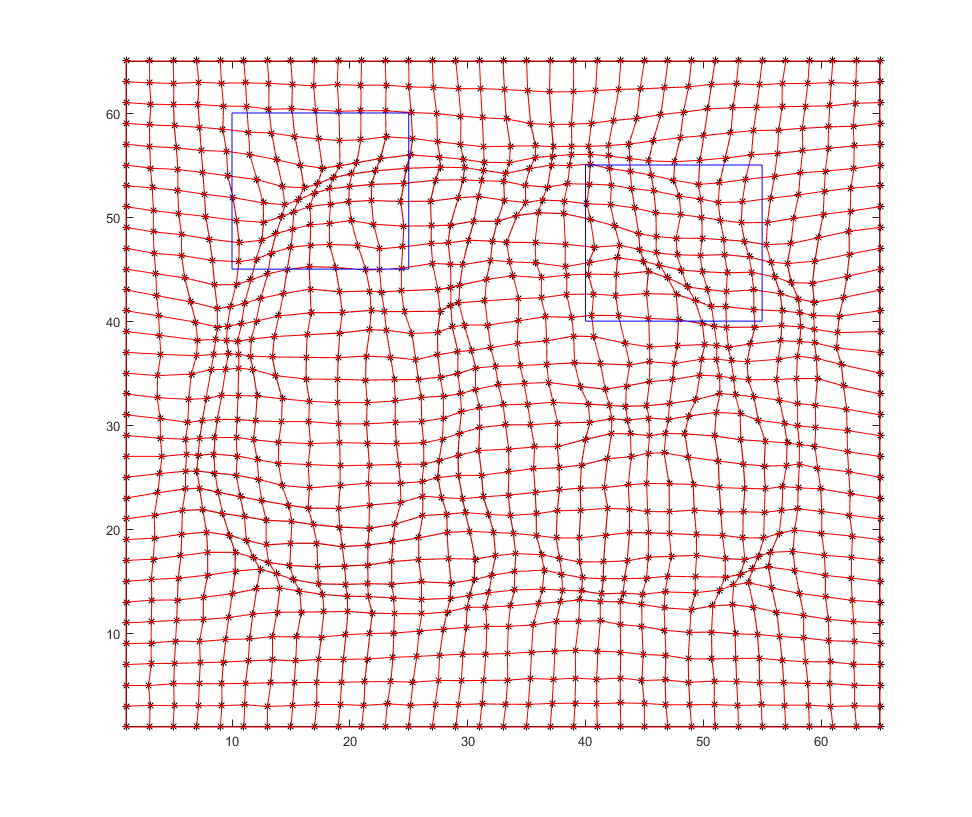}}}\\
	\subfigure[Enlarged view of the left rectangle]{{\includegraphics[width=6cm,height=5.4cm]{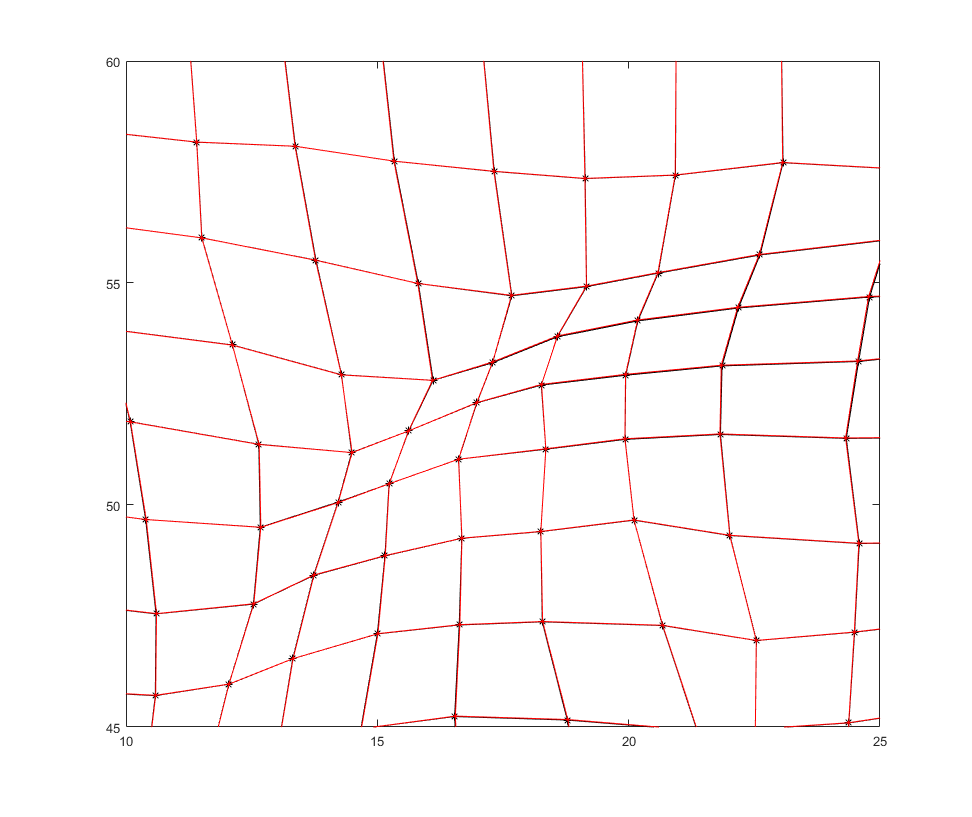}}}
	\subfigure[Enlarged view of the right rectangle]{{\includegraphics[width=6cm,height=5.4cm]{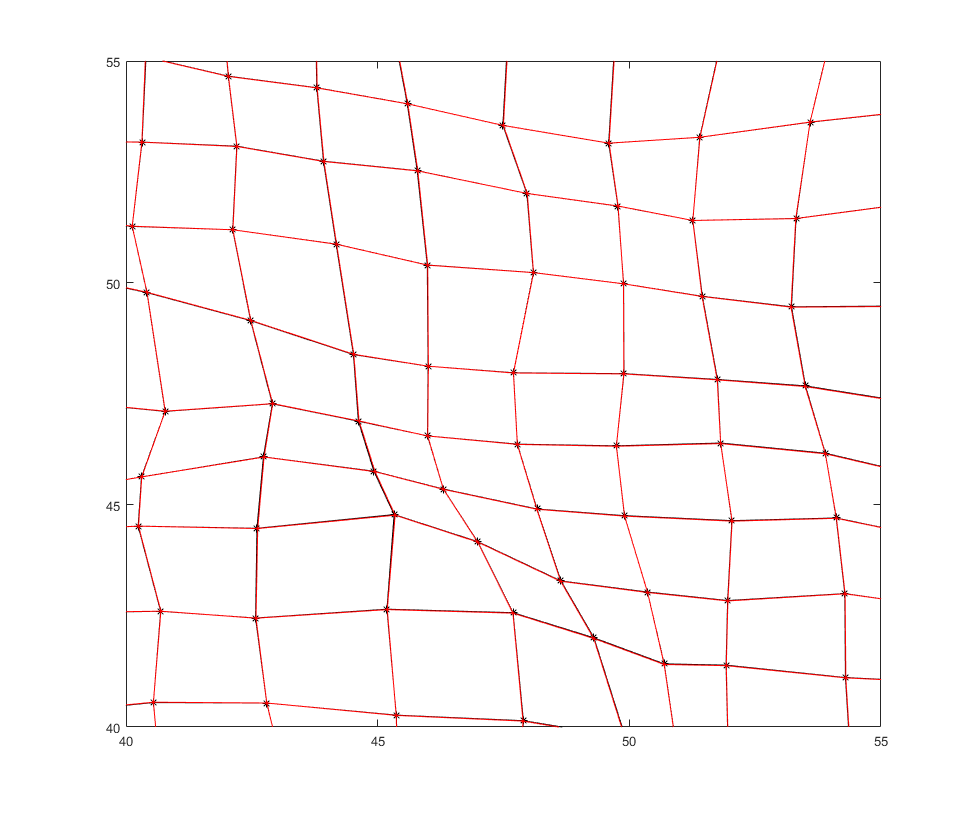}}}
\end{figure}
As it can be seen in Figures (d), (e), and (f), that $\pmb{T_{2}}$ has no obvious visual difference from $\pmb{T_{0}}$. This example suggests that the transformation $\pmb{T_{0}}$ is uniquely determined by its Jacobian determinant, $\text{det}\nabla(\pmb{T_{0}})$, and its curl-vector, $\text{curl}(\pmb{T_{0}})$.}

Inspired by these numerical results, it is conjectured that a desired transformation can be uniquely determined by the prescribed positive Jacobian determinant together with the curl-vector.

In section 2, the $variational$ $method$ is briefly reviewed, which achieves both the prescribed positive Jacobian determinant and the curl-vector in $L^{2}-norm$. In section 3, we outline the steps to prove a version of the uniqueness problem, which is based on an iterative procedure on a Sobolev's space. The $Green's$ formula and the $Poincare's$ inequality are used in the key components of this approach.  
\section{The $Variational$ $Method$ for Mesh Generation}
Let $\mathrm{\Omega} \subset \mathbb R^n$, $n=2$, or $3$ be the domain and a scalar function on $f_0(\pmb{x})>0$ and a vector-valued $\pmb{g}_0(\pmb{x})$ on the domain $\mathrm{\Omega}$ with
	\begin{equation}
	\int_{\mathrm{\Omega}} f_0(\pmb{x})d\pmb{x} = |\mathrm{\Omega}| \text{ and } \nabla \cdot  \pmb{g}_0(\pmb{x})= 0
	\end{equation}
We look for a diffeomorphism 
	\begin{equation} 
	\pmb{\phi}(\pmb{x}):\mathrm{\Omega}\rightarrow\mathrm{\Omega}
	\end{equation}
	\begin{equation} 
	\pmb{\phi}(\pmb{x})=\pmb{\phi}_{old}(\pmb{x})+\pmb{u}(\pmb{x})
	\end{equation}
that minimizes the cost functional --- sum of squared difference:
	\begin{equation}
	ssd = \dfrac{1}{2}\int_{\mathrm{\Omega}} [(J(\pmb{\phi}(\pmb{x})) - f_0(\pmb{x}))^2+|\text{curl}(\pmb{\phi}(\pmb{x}))- \pmb{g}_0(\pmb{x})|^2] d\pmb{x}
	\end{equation}
where $J(\pmb{\phi}(\pmb{x})) =\text{det}\nabla(\pmb{\phi}(\pmb{x}))$, subject to the following constraints with control functions $f(\pmb{x})$ and $\pmb{g}(\pmb{x})$
\begin{equation*}
\left\{
	\begin{aligned}
	\text{div } \pmb{u}(\pmb{x})& = f(\pmb{x}) \qquad in\quad \mathrm{\Omega}\\
	\text{curl } \pmb{u}(\pmb{x})& = \pmb{g}(\pmb{x}) \qquad in \quad \mathrm{\Omega}\\
	\pmb{u}(\pmb{x})& = \pmb{0} \qquad  \text{ on } \partial\mathrm{\Omega}
	\end{aligned}\right.
\end{equation*}
The above div-curl equations lead to	
\begin{equation*}
\left\{
	\begin{aligned}
	\mathrm{\Delta} u_1& =f(\pmb{x})_{x_{1}}-g_{3x_{2}}(\pmb{x})+g_{2x_{2}}(\pmb{x})=F_1(\pmb{x})\\
	\mathrm{\Delta} u_2& =f(\pmb{x})_{x_{2}}-g_{3x_{1}}(\pmb{x})-g_{1x_{3}}(\pmb{x})=F_2(\pmb{x})\\
	\mathrm{\Delta} u_3& =f(\pmb{x})_{x_{3}}-g_{2x_{1}}(\pmb{x})+g_{1x_{2}}(\pmb{x})=F_3(\pmb{x})\\
	\end{aligned}\right.
\end{equation*}
i.e.,
\begin{equation}\label{eq25}
\left\{
	\begin{aligned}
	\mathrm{\Delta} \pmb{u}(\pmb{x}) &= \pmb{F}(\pmb{x}) \qquad \text{ in } \mathrm{\Omega}\\
	\pmb{u}(\pmb{x})& = \pmb{0} \qquad  \text{ on } \partial\mathrm{\Omega}
	\end{aligned}\right.
\end{equation}	

Since both of the theoretical derivations of the $variational$ $method$ and more of its numerical examples had been included in \cite{XiChen}\cite{XiChen2}, we go straight to the proposed approach to the uniqueness conjecture.

\section{The Uniqueness Problem}
Suppose two smooth transformations $\pmb{\phi}$, $\pmb{\psi}: \mathrm{\Omega}\rightarrow\mathrm{\Omega}$, $\mathrm{\Omega} \subset \mathbb R^3$  have the same Jacobian determinant and curl-vector, namely, 
\begin{equation}\label{eq26}
  J(\pmb{\phi})=J(\pmb{\psi})
\end{equation}  
\begin{equation}\label{eq27}
  \text{curl}(\pmb{\phi})=\text{curl }(\pmb{\psi})
\end{equation}  
\begin{equation}\label{eq28}
  \pmb{\phi}=\pmb{\psi} \text{, on } \partial\mathrm{\Omega}
\end{equation}
where $J(\pmb{\phi}) =\text{det}\nabla(\pmb{\phi})$. We would like to ask: can the above conditions guarantee that $\pmb{\phi}\equiv\pmb{\psi} \text{ on } \mathrm{\Omega}$?

$\mathbf{First}$, we consider a simple case as follows. Let $\pmb{\phi}$ and $\pmb{\psi}: \mathrm{\Omega}\rightarrow\mathrm{\Omega}$ be two smooth transformations by
\begin{equation} \label{eq29}
  \pmb{\phi}=\pmb{id}+\pmb{u}
\end{equation}
\begin{equation}\label{eq210}
  \pmb{\psi}=\pmb{id}
\end{equation}
where $\pmb{u}$ is sufficiently small transformation in the Sobolev space ${H_{0}^{2}(\mathrm{\Omega})}$. One can suppose that $\pmb{u}$ satisfies
\begin{equation}\label{eq211}
	 \left\|\pmb{u}\right\|_{H_{0}^{2}(\mathrm{\Omega})} < \epsilon
\end{equation}
Hence, we have 
\begin{equation}\label{eq212}
\left\{
	\begin{aligned}
	\left\|\pmb{u}\right\|_{L^{2}}& <\epsilon\\
	\left\|\nabla \pmb{u}\right\|_{L^{2}}&< \epsilon\\
	\left\|\mathrm{\Delta}\pmb{u}\right\|_{L^{2}}&< \epsilon\\
	\end{aligned}\right.
\end{equation} 

$\mathbf{Second}$, from $\pmb{\phi}(x_1,x_2,x_3)=(x_1+u_1(x_1,x_2,x_3),x_2+u_2(x_1,x_2,x_3),x_3+u_3(x_1,x_2,x_3))$ and $\pmb{\psi}(x_1,x_2,x_3)=(x_1,x_2,x_3)$
, we may derive 
\begin{equation}
J(\pmb{\phi})=\begin{vmatrix}
1+u_{1x_{1}} & u_{2x_{1}} & u_{3x_{1}}\\ 
u_{1x_{2}} & 1+u_{2x_{2}} & u_{3x_{2}}\\ 
u_{1x_{3}} & u_{2x_{3}} & 1+u_{3x_{3}}\notag   
\end{vmatrix}
\end{equation}
\begin{equation*}
=1+u_{1x_{1}}+u_{2x_{2}}+u_{3x_{3}}
\end{equation*}
\begin{equation*}
+ u_{1x_{1}}u_{2x_{2}}u_{3x_{3}}+ u_{1x_{3}}u_{2x_{1}}u_{3x_{2}}+ u_{1x_{2}}u_{2x_{3}}u_{3x_{1}}
\end{equation*}
\begin{equation*}
- u_{1x_{1}}u_{2x_{3}}u_{3x_{2}}- u_{1x_{2}}u_{2x_{1}}u_{3x_{3}}- u_{1x_{3}}u_{2x_{2}}u_{3x_{1}}
\end{equation*}
\begin{equation*}
+ u_{1x_{1}}u_{2x_{2}}+ u_{1x_{1}}u_{3x_{3}}+ u_{2x_{2}}u_{3x_{3}}
\end{equation*}
\begin{equation*}
- u_{1x_{2}}u_{2x_{1}}- u_{1x_{3}}u_{3x_{1}}- u_{2x_{3}}u_{3x_{2}}
\end{equation*}
\begin{equation*}
=1+\text{div}(\pmb{u})+J(\pmb{u})
\end{equation*}
\begin{equation*}
+ u_{1x_{1}}u_{2x_{2}}+ u_{1x_{1}}u_{3x_{3}}+ u_{2x_{2}}u_{3x_{3}}
\end{equation*}
\begin{equation*}
- u_{1x_{2}}u_{2x_{1}}- u_{1x_{3}}u_{3x_{1}}- u_{2x_{3}}u_{3x_{2}}
\end{equation*}
\begin{equation*}
=1+\text{div}(\pmb{u})-\mathcal{F}(\pmb{u})
\end{equation*}
where we denote
\begin{equation*}
\mathcal{F}(\pmb{u})=-[J(\pmb{u})+Tail(\pmb{
u})]
\end{equation*} and
\begin{equation*}
Tail(\pmb{
u})= u_{1x_{1}}u_{2x_{2}}+ u_{1x_{1}}u_{3x_{3}}+ u_{2x_{2}}u_{3x_{3}}
\end{equation*} 
\begin{equation}\label{eq213}
- u_{1x_{2}}u_{2x_{1}}- u_{1x_{3}}u_{3x_{1}}- u_{2x_{3}}u_{3x_{2}}
\end{equation}
According to (\ref{eq26}) and (\ref{eq27}) we have,
\begin{equation*}
	0=J(\pmb{\phi})-J(\pmb{\psi})= 1+div(\pmb{u})- \mathcal{F}(\pmb{u})-1=div(\pmb{u})-\mathcal{F}(\pmb{u}) 
\end{equation*} 
\begin{equation*}
	\pmb{0}=\text{curl}(\pmb{\phi})-\text{curl}(\pmb{\psi})=
	\begin{pmatrix}
	u_{3x_{2}}-u_{2x_{3}}\\ 
	u_{1x_{3}}-u_{3x_{1}}\\ 
	u_{1x_{2}}-u_{2x_{1}} \notag 
	\end{pmatrix}-\pmb{0}= \text{curl }(\pmb{u}) 
\end{equation*}
i.e.,
\begin{equation}\label{eq214}
\left\{
	\begin{aligned}
	\text{div}(\pmb{u})& =\mathcal{F}(\pmb{u})\\
	\text{curl}(\pmb{u})& = \pmb{0}.
	\end{aligned}\right.
\end{equation}
It follows from (\ref{eq214}) that $\pmb{u}$ satisfies the $Poisson$ equations:
\begin{equation*}
\left\{
	\begin{aligned}
	\mathrm{\Delta} u_1& =\text{div}(\pmb{u})_{x_{1}}-[\text{curl}(\pmb{u})_{3}]_{x_{2}}+[\text{curl}(\pmb{u})_{2}]_{x_{3}}=\mathcal{F}(\pmb{u})_{x_{1}}\\
	\mathrm{\Delta} u_2& =\text{div}(\pmb{u})_{x_{2}}-[\text{curl}(\pmb{u})_{3}]_{x_{1}}-[\text{curl}(\pmb{u})_{1}]_{x_{3}}=\mathcal{F}(\pmb{u})_{x_{2}}\\
	\mathrm{\Delta} u_3& =\text{div}(\pmb{u})_{x_{3}}-[\text{curl}(\pmb{u})_{2}]_{x_{1}}+[\text{curl}(\pmb{u})_{1}]_{x_{2}}=\mathcal{F}(\pmb{u})_{x_{3}}\\
	\end{aligned}\right.
\end{equation*}
\begin{equation*}
	\Rightarrow \mathrm{\Delta} \pmb{u} =	(\mathcal{F}(\pmb{u})_{x_{1}},\mathcal{F}(\pmb{u})_{x_{2}},\mathcal{F}(\pmb{u})_{x_{3}})= {\nabla_{\pmb{x}}\mathcal{F}(\pmb{u})}
\end{equation*}
Note that the dominating terms of ${\mathcal{F}(\pmb{u})}$ at (\ref{eq213}) are products of only the first partial derivatives of $\pmb{u}$, so  the dominating terms of $\nabla_{\pmb{x}}{\mathcal{F}(\pmb{u})}$ at (\ref{eq215}) are the terms in products of the first and second partial derivatives of $\pmb{u}$. This means, by (\ref{eq212}), we get 
\begin{equation}\label{eq215}
\left\|\mathrm{\Delta}\pmb{u} \right\|_{L^{2}}=(\int_{\mathrm{\Omega}}|\mathrm{\Delta} \pmb{u}|^2)^{\frac{1}{2}}=(\int_{\mathrm{\Omega}}| \nabla_{\pmb{x}}{\mathcal{F}(\pmb{u})}|^2)^{\frac{1}{2}}=\left\|\nabla_{\pmb{x}}{\mathcal{F}(\pmb{u})} \right\|_{L^{2}}
\end{equation}
\begin{equation}\label{eq216}
\Rightarrow\left\|\mathrm{\Delta}\pmb{u} \right\|_{L^{2}}<\epsilon\cdot\epsilon=\epsilon^{2}
\end{equation}
Next, we will establish an inequality for $\left\|\pmb{u}\right\|_{L^{2}}$. Since $\pmb{u}=0\text{ on } \partial\mathrm{\Omega}$, by $Green's$ formula, we can derive,
\begin{equation}\label{eq217}
\int_{\mathrm{\Omega}}|\nabla \pmb{u}|^2=|\int_{\mathrm{\Omega}}\pmb{u} \cdot \mathrm{\Delta}\pmb{u}|
\end{equation}
Applying the $Cauchy-Schwarz$ inequality and properties of integration to the RHS of (\ref{eq217}) to get
\begin{equation*}
\int_{\mathrm{\Omega}}|\nabla \pmb{u}|^2=|\int_{\mathrm{\Omega}}\pmb{u} \cdot \mathrm{\Delta}\pmb{u}|\leq\int_{\mathrm{\Omega}}|\pmb{u} ||\mathrm{\Delta}\pmb{u}|\leq\left\|\pmb{u}\right\|_{L^{2}}\left\|\mathrm{\Delta}\pmb{u}\right\|_{L^{2}}
\end{equation*}
\begin{equation}\label{eq218}
\Rightarrow \left\|\nabla \pmb{u}\right\|_{L^{2}}^{2}\leq\left\|\pmb{u}\right\|_{L^{2}}\left\|\mathrm{\Delta}\pmb{u}\right\|_{L^{2}}
\end{equation}
Applying the $Poincare's$ inequality to the LHS of (\ref{eq217}), $\exists$ $0<C\in\mathbb{R}$, such that
\begin{equation}\label{eq219}
\left\|\pmb{u}\right\|_{L^{2}}^{2}\leq C \left\|\nabla \pmb{u}\right\|_{L^{2}}^{2}= C \int_{\mathrm{\Omega}}|\nabla \pmb{u}|^2
\end{equation}
Next, we combine (\ref{eq218}) and (\ref{eq219}) to have
\begin{equation}\label{eq220}
\left\|\pmb{u}\right\|_{L^{2}}^{2}\leq C \left\|\nabla \pmb{u}\right\|_{L^{2}}^{2}\leq C \left\|\pmb{u}\right\|_{L^{2}}\left\|\mathrm{\Delta}\pmb{u}\right\|_{L^{2}}
\end{equation}
\begin{equation}\label{eq221}
\Rightarrow \left\|\pmb{u}\right\|_{L^{2}}\leq C \left\|\mathrm{\Delta}\pmb{u}\right\|_{L^{2}}< C \epsilon^2
\end{equation}
And, as for $\left\|\mathrm{\Delta}\pmb{u}\right\|_{L^{2}}$, by (\ref{eq216}), (\ref{eq218}) and (\ref{eq221}), we may bound $\left\|\nabla \pmb{u}\right\|_{L^{2}}$ as 
\begin{equation}\label{eq222}
\left\|\nabla \pmb{u}\right\|_{L^{2}}\leq(\left\|\pmb{u}\right\|_{L^{2}}\left\|\mathrm{\Delta}\pmb{u}\right\|_{L^{2}})^{\frac{1}{2}}
\end{equation}
\begin{equation}\label{eq223}
\Rightarrow  \left\|\nabla \pmb{u}\right\|_{L^{2}}<(C \epsilon^2 \cdot\epsilon^{2})^{\frac{1}{2}}=C^{\frac{1}{2}}\epsilon^{2}
\end{equation}
Now, $\pmb{u}$ satisfies 
\begin{equation}\label{eq224}
\left\{
	\begin{aligned}
	\left\|\pmb{u}\right\|_{L^{2}}& <C\epsilon^2\\
	\left\|\nabla \pmb{u}\right\|_{L^{2}}& < C^{\frac{1}{2}}\epsilon^2\\
	\left\|\mathrm{\Delta}\pmb{u}\right\|_{L^{2}}& < \epsilon^2\\
	\end{aligned}\right.
\end{equation}

$\mathbf{Third}$, let's treat the procedure of (\ref{eq212}) to (\ref{eq224}) as Step-0, and repeat it again with replacing (\ref{eq212}) by the result (\ref{eq224}). So, plug (\ref{eq223}) into the first derivative of $\pmb{u}$ in (\ref{eq215}), we get 
\begin{equation}\label{eq225}
\left\|\mathrm{\Delta}\pmb{u}\right\|_{L^{2}} =\left\|\nabla_{\pmb{x}}{\mathcal{F}(\pmb{u})} \right\|_{L^{2}}<\epsilon\cdot C^{\frac{1}{2}}\epsilon^2=C^{\frac{1}{2}}\epsilon^3
\end{equation}
Then, by (\ref{eq220}), we get
\begin{equation}\label{eq226}
\left\|\pmb{u}\right\|_{L^{2}}\leq C \left\|\mathrm{\Delta}\pmb{u}\right\|_{L^{2}}< C \cdot C^{\frac{1}{2}} \epsilon^3=C^{(1+\frac{1}{2})}\epsilon^3
\end{equation}
and by (\ref{eq222}), we get
\begin{equation}\label{eq227}
\left\|\nabla \pmb{u}\right\|_{L^{2}}\leq(\left\|\pmb{u}\right\|_{L^{2}}\left\|\mathrm{\Delta}\pmb{u}\right\|_{L^{2}})^{\frac{1}{2}}<(C^{(1+\frac{1}{2})} \epsilon^3 \cdot C^{\frac{1}{2}}\epsilon^{3})^{\frac{1}{2}}=C\epsilon^{3}
\end{equation}
which above can be combined from (\ref{eq224}), (\ref{eq225}) and (\ref{eq226}) into 
\begin{equation}\label{eq228}
\left\{
	\begin{aligned}
	\left\|\pmb{u}\right\|_{L^{2}}& <C^{(1+\frac{1}{2})}\epsilon^3\\
	\left\|\nabla \pmb{u}\right\|_{L^{2}}& < C\epsilon^3\\
	\left\|\mathrm{\Delta}\pmb{u}\right\|_{L^{2}}& < C^{\frac{1}{2}}\epsilon^3\\
	\end{aligned}\right.
\end{equation}

$\mathbf{Fourth}$, in order to complete the argument of the simple case, an iterative process based on the procedure (\ref{eq224}) to (\ref{eq228}) is constructed as follows:  
\begin{algorithm}[H]
	\begin{itemize}
		\item Step-0: From (\ref{eq212}) to (\ref{eq224}), and
		 denote (\ref{eq224}) as $(\ref{eq224})_{k}$, set $k=0$
		
		\item Step-1: Start iteration (Step-2 to 5) on $k=k+1$
		
		\item Step-2: Apply $(\ref{eq224})_{k}$ on (\ref{eq215}) to get $\left\|\mathrm{\Delta}\pmb{u} \right\|_{L^{2}} <C^{(0+\frac{k}{2})}\epsilon^{(2+k)}$, and denote it as $(\ref{eq215})_{k}$
		
		\item Step-3: Apply $(\ref{eq225})_{k}$ on (\ref{eq220}) to get $\left\|\pmb{u}\right\|_{L^{2}} <C^{(1+\frac{k}{2})}\epsilon^{(2+k)}$, and denote it as $(\ref{eq220})_{k}$
		
		\item Step-4: Apply $(\ref{eq220})_{k}$ on (\ref{eq222}) to get $\left\|\nabla \pmb{u}\right\|_{L^{2}}<C^{(\frac{1}{2}+\frac{k}{2})}\epsilon^{(2+k)}$, and denote it as $(\ref{eq222})_{k}$
		
		\item Step-5: Combine $(\ref{eq215})_{k}$, $(\ref{eq220})_{k}$, $(\ref{eq222})_{k}$ to form $(\ref{eq224})_{k+1}$ (For example: $(\ref{eq224})_{1}$ is (\ref{eq228})), then back to Step-1
	\end{itemize}
	\hrule 
\end{algorithm}
Hence, on the $k$-th iteration, (\ref{eq224}) can be improved to $(\ref{eq224})_{k}$, i.e., 
\begin{equation*}\label{eq224k}
\left\{
	\begin{aligned}
	\left\|\pmb{u}\right\|_{L^{2}}& <C^{(1+\frac{k}{2})}\epsilon^{(2+k)}\\
	\left\|\nabla \pmb{u}\right\|_{L^{2}}& < C^{(\frac{1}{2}+\frac{k}{2})}\epsilon^{(2+k)}\\
	\left\|\mathrm{\Delta}\pmb{u} \right\|_{L^{2}}& <C^{(0+\frac{k}{2})}\epsilon^{(2+k)}\\
	\end{aligned}\right.
\end{equation*}
It is natural to take a step inductively forward at $k+1$. So, plug $(\ref{eq224})_{k}$ into (\ref{eq215}), we get
\begin{equation*}
\left\|\mathrm{\Delta}\pmb{u} \right\|_{L^{2}}=\left\|\nabla_{\pmb{x}}{\mathcal{F}(\pmb{u})} \right\|_{L^{2}}<\epsilon \cdot C^{(\frac{1}{2}+\frac{k}{2})}\epsilon^{(2+k)}=C^{(0+\frac{k+1}{2})}\epsilon^{(2+k+1)}
\end{equation*}
and denote it as $(\ref{eq215})_{k+1}$. Then, by (\ref{eq220}) we get 
\begin{equation*}
\left\|\pmb{u}\right\|_{L^{2}}\leq C \left\|\mathrm{\Delta}\pmb{u}\right\|_{L^{2}}< C \cdot C^{(0+\frac{k+1}{2})}\epsilon^{(2+k+1)}=C^{(1+\frac{k+1}{2})}\epsilon^{(2+k+1)}
\end{equation*}
and denote it as $(\ref{eq220})_{k+1}$. Then, by (\ref{eq222}) we get
\begin{equation*}
\left\|\nabla \pmb{u}\right\|_{L^{2}}<(C^{(1+\frac{k+1}{2})} \epsilon^{(2+k+1)} \cdot C^{(0+\frac{k+1}{2})}\epsilon^{(2+k+1)})^{\frac{1}{2}}=C^{(\frac{1}{2}+\frac{k+1}{2})}\epsilon^{(2+k+1)}
\end{equation*}
and denote it as $(\ref{eq222})_{k+1}$. Which above leads to the immediate inductive step $(\ref{eq224})_{k+1}$
\begin{equation*}\label{eq224k+1}
\left\{
	\begin{aligned}
	\left\|\pmb{u}\right\|_{L^{2}}& <C^{(1+\frac{k+1}{2})}\epsilon^{(2+k+1)}\\
	\left\|\nabla \pmb{u}\right\|_{L^{2}}& < C^{(\frac{1}{2}+\frac{k+1}{2})}\epsilon^{(2+k+1)}\\
	\left\|\mathrm{\Delta}\pmb{u} \right\|_{L^{2}}& <C^{(0+\frac{k+1}{2})}\epsilon^{(2+k+1)}\\
	\end{aligned}\right.
\end{equation*}

Therefore, the system of inequalities $(\ref{eq222})_{k}$ is iterated to reduce the bound of $\left\|\pmb{u}\right\|_{L^{2}}$ for every increment of $k=k+1$.
Since $\pmb{u}$ satisfies (\ref{eq212}), then we may conclude such $\pmb{u}\text{ on } \mathrm{\Omega}$ is also satisfying $\left\|\pmb{u}\right\|_{L^{2}} <C^{(1+\frac{k}{2})}\epsilon^{(2+k)}$, where $C^{(1+\frac{k}{2})}\epsilon^{(2+k)}$ converges to $0$ by the choice of $0<\epsilon<\min\{1, 1/\sqrt{C}\}$, as $ k \longrightarrow \infty$. So it can also be concluded that $\left\|\pmb{\phi}-\pmb{\psi} \right\|_{L^{2}}=\left\|\pmb{u} \right\|_{L^{2}}\longrightarrow 0$ as $ k \longrightarrow \infty$, therefore 
$\pmb{\phi} \equiv \pmb{\psi}=\pmb{id}$ on $\mathrm{\Omega}$.

\section*{Conclusion}
In this research note, we describe an approach to the uniqueness problem based on the simple case which the two smooth transformations are close to each other and one of them is the $\pmb{identity}$ map. The general uniqueness problem from (\ref{eq26}) to (\ref{eq28}) remains open. An interesting intermediate step is to show that, for any two sufficiently close $\pmb{\phi}$, $\pmb{\psi}$, a similar argument can be applied. 

%


\begin{thebibliography}{12}
%
\bibitem {Ands}
Anderson, D. and Liao, G.:
An New Approach to  Grid Generation. 
Applicable Analysis: An international Journal, Vol. 44, no. 3-4, pp. 285-298 (1992)
%
\bibitem {Cai}
Cai, X. X., Fleitas, D., Jiang, B. and Liao, G.:
Adaptive grid generation based on the least-squares finite elements method.
Computers and Mathematics with Applications, vol. 48, no 7-8, pp. 1007-1085 (2004)
%
\bibitem {XiChen}
Chen, X.:
Numerical Construction of Diffeomorphism and the Applications to Grid Generation and Image Registration. Dissertation, University of Texas at Arlington (2015)
%
\bibitem {XiChen2}
Chen, X. and Liao, G.:
New Variational Method of Grid Generation with Prescribed Jacobian determinant and Prescribed Curl.
arxiv.org/pdf/1507.03715 (2015)
%
\bibitem {Dac:Mos}
Dacoragna, B. and Moser, J.:
On A Partial Differential Equation Involving the Jacobian determinant.
Ann. Inst H Poincaré, vol. 7, no. 1, pp. 1-26 (1990)
%
\bibitem {Liao}
Liao, G, Cai, X. X., Liu, J, Luo, X, Wang, J and Xue, J:
Construction of Differentiable Transformations.
Applied Mathematics Letters, Vol. 22, pp. 1543-1548 (2009)%
%
\bibitem {Liu}
Liu, F., Ji, S. and Liao, G.:
An Adaptive Grid method and its Application to Steady Euler Flow calculations. 
SIAM J. Sci. Comput, Vol. 20, no. 3 pp. 811-825 (1998)
%
\bibitem {Zhou}
Zhou, Z., Chen, X and Liao, G.:
A Novel Deformation Method for Higher Order Mesh Generation.
arxiv.org/abs/1710.00291 (2017)
%

\end{thebibliography}
\end{document}